# The 150th Anniversary of Fusakichi Omori

**Anatol V. Guglielmi, Alexey D. Zavyalov**

*Schmidt Institute of Physics of the Earth, Russian Academy of Sciences, ul. B. Gruzinskaya 10, Moscow, 123995 Russia*
*e-mail: guglielmi@mail.ru, zavyalov@ifz.ru*

**Abstract**

This paper is devoted to the memory of the outstanding Japanese scientist. In 1896, Fusakichi Omori discovered the law of the aftershocks evolution that bears his name. We represent the Omori law in the form of a differential equation. This allows us to take into account the non-stationarity of rocks in the earthquake source, which "cools down" after the main shock.

**Contents**

## 1. Introduction

The outstanding Japanese seismologist Fusakichi Omori was born 150 years ago, October 30, 1868 [Davison, 1927]. He deserves credit for the discovery of the first law of earthquake physics. The Omori law states that after strong earthquake, the frequency of aftershocks $n$, i.e., the repeated shocks that follow the main shock, decay with time $t$, on average, according to the hyperbolic law

$$n = k/(c+t). \qquad (1)$$

Here $k > 0$, $c > 0$, and $t \geq 0$ [Omori, 1894].

It was really the first law in the field of earthquake physics, if we talk about the chronological sequence of outstanding discoveries in seismology. As the second law, the Gutenberg-Richter law describing the distribution of earthquakes over magnitudes must be mentioned [Gutenberg, Richter, 1944]. A brief description of five laws of seismology can be found in [Costa et al., 2016]. We will not dwell on this in order to not be diverted from our topic.

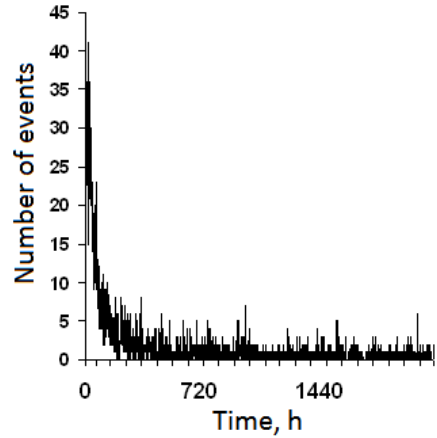

Fig. 1. Time dependence of the number of aftershocks, 24.11.1984, California.

It is appropriate here to bring up a typical example of aftershock evolution. Figure 1 shows an aftershock sequence after the main shock with the magnitude $M = 6.6$ that took place on 24.11.1984 at 13 h 15 min 56 s universal time. The main shock epicenter was located in Southern California. The hypocenter was at a depth of 10 km, and 3663 aftershocks occurred in the zone $0.5º$ in radius during the 90 days after the main shock. We see that on average the frequency of aftershocks decay hyperbolically with time.

Professor Omori has lived the active life, full of creative searches. He traveled a lot, gained international fame, and was the president of the Japanese Imperial Earthquake Investigation Committee. The circumstances of his death were tragic [Davison, 1924]. Great Kanto earthquake occurred on Saturday, September 1, 1923. Yokohama and Tokyo were destroyed. Many people died. At this time, Fusakichi Omori was in Australia. Upon learning of the disaster, he left for home immediately. During the voyage he fell seriously ill and died on November 8, 1923 at the age of 55 years, shortly after his return to Tokyo.

Our paper is devoted to the memory of the outstanding scientist. In the next section we represent the Omori law in the form of a differential equation. This allows us to take into account the non-stationarity of rocks in the earthquake source, which "cools down" after the main shock.

## 2. Aftershocks equation

Let us represent the Omori law (1) in the form of a differential equation

$$dn/dt + \sigma n^2 = 0, \qquad (2)$$

in which $\sigma = k^{-1}$. We will call it the equation of the evolution of aftershocks, and $\sigma$ the deactivation coefficient of the earthquake focus [Guglielmi, 2016]. At first glance, Eq. (2) is just one more representation of the hyperbolic Omori law (1), but only at first glance. We must take into account that in modeling natural phenomena, it is easier in many cases to work with the evolution equation than with the set of its solutions. In our case, the representation of formula (1) in the form of equation (2) opens an interesting possibility to take into account the non-stationarity of rocks in the earthquake source at the phenomenological level. To do this, it suffices to take into account the time dependence of $\sigma$ in Eq. (2). Then the evolution equation will have the form

$$n(t) = n_0 \left[1 + n_0 \int_0^t \sigma(t') dt'\right]^{-1}. \qquad (3)$$

Here $n_0$ is the frequency of aftershocks at the moment $t=0$. The distinction between (3) and (1) consists only in the fact that the time is replaced by the integral of deactivation coefficient over time. If and only if $\sigma = \text{const}$, then (3) coincides with (1), with $k = 1/\sigma$, and $c = 1/n_0\sigma$.

It is interesting to compare our interpretation Omori law with the two-parameter modification of the law used in the papers [Hirano, 1924; Utsu, 1961] and widely known in seismology:

$$n = k/(c+t)^p. \qquad (4)$$

The parameter *p* varies from case to case in a wide range, from about 0.5 to 1.5, with the mean noticeable larger than unity: p = 1.1 (see, for example, the review papers [Utsu et al., 1995; Guglielmi, 2017] and the literature cited therein).

Let us assume for a moment that the Hirano-Utsu law (4) is satisfied. Then

$$\sigma = k^{-1} p (c+t)^{p-1}. \qquad (5)$$

We see that for *p* < 1 (*p* > 1), the deactivation coefficient monotonically decreases (increases) over time. In other words, the deviation of the parameter *p* from unity observed experimentally indicates the non-stationarity of the medium. When $p \neq 1$, the deactivation coefficient $\sigma$ varies over time as a result of relaxation processes occurring in the earthquake source. Figuratively speaking, the focus gradually "cools down" after the main shock.

To verify the relationship (5) we made a pilot analysis of the ten series of aftershocks. The deactivation factor was calculated by the formula $\sigma = -n^{-1} d \ln n / dt$. In none of the ten series the relation (5) was confirmed. (A detailed report will be submitted in a separate paper.) Moreover, in some cases, there was a non-monotonic dependence of $\sigma$ on time, which clearly demonstrates the impermanence of the parameter *p* within the same series of aftershocks. We recall in this connection that the variability of *p* within the series was also noted in the work [Hirano, 1924]. However, it is quite obvious that it is undesirable to approximate the real stream of aftershocks using formula (4), in which the parameter $p$ depends on time. In this sense, the formula (3) is more preferable than (4). We draw attention to the fact that in (3) the hyperbolicity of the law is preserved. However, unlike (1), in (3) it is taken into account that the time in the source, figuratively speaking, is flowing unevenly.

## 3. Discussion

We would like to discuss the question of how to use the aftershocks equation for setting the inverse problem of the physics of earthquake source. Let us consider a linear Volterra integral equation of the first kind

$$\int_0^t K(t,t')\sigma(t')\,dt' = g(t). \tag{6}$$

Here $\sigma(t)$ is an unknown function, and $g(t)=[n_0 n(t)]^{-1}[n_0 - n(t)]$ is the known function. The inverse problem is to determine the deactivation coefficient $\sigma(t)$ from the observation data of the frequency of aftershocks $n(t)$. The choice of kernel $K(t,t')$ should be the subject of further study of the problem. But even if we use the trivial kernel $K=1$, then in this simplest case the inverse problem still has a physical sense. Indeed, if $K=1$, then equation (6) follows directly from equation (2). In this case $\sigma = dg/dt$.

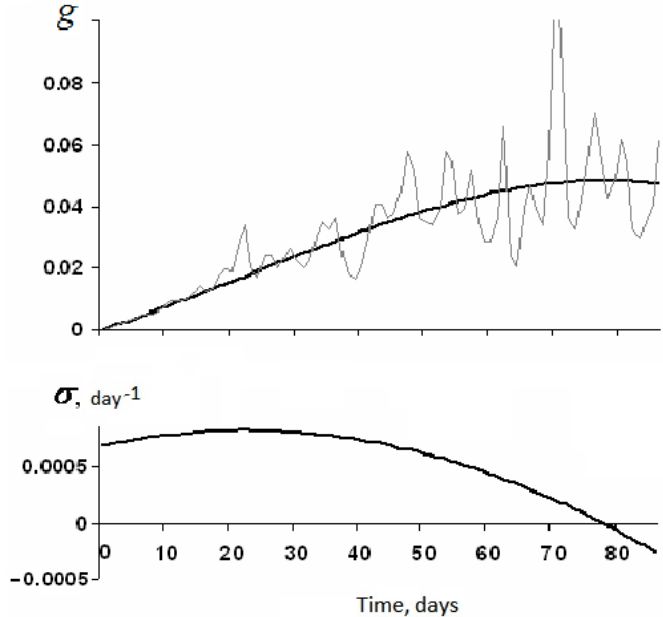

Fig. 2. Results of analyzing the aftershocks shown in Fig. 1. Top panel: The function $g(t)$ before and after smoothing (the gray and black lines respectively). Bottom panel: The time dependence of the deactivation coefficient $\sigma$.

We consider two examples of solving the inverse problem in this formulation. Let us take the event shown in Fig. (1). The experimental data on the dependence $g(t)$ should be approximated by a sufficiently smooth function before calculating the deactivation coefficient by the formula $\sigma = dg/dt$. The result is shown in Fig. 2 (bottom panel). We see that the quasi-stationarity of the process is presented on average for approximately 45 days after the main shock. Within this time interval (it seems natural to call it the Omori epoch), we have approximately $\sigma = 7.8 \cdot 10^{-4}$ days$^{-1}$ or $k = 1/\sigma \approx 1300$ days. Then $\sigma$ starts to decrease and decays to zero, whereas fluctuations in $g$ grow rapidly (see the gray line in Fig. 2, top panel).

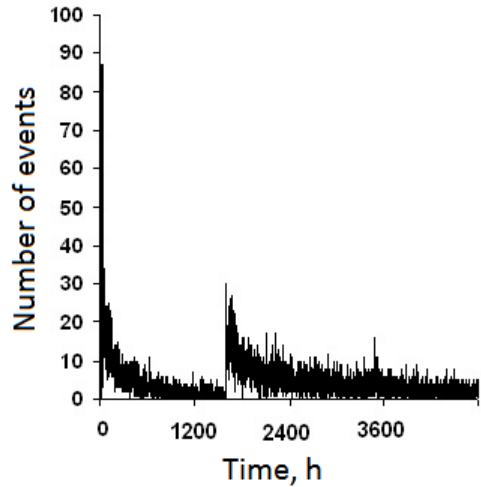

Fig. 3. Time dependence of the number of aftershocks after two main shocks with magnitudes $M = 6.1$ and $M = 7.3$ that respectively occurred on 23.04.1992 and 28.06.1992, in California.

We consider one more event shown in Fig. 3. It is of interest as an example of a doublet of main shocks. The first shock ($M = 6.1$) occurred on 23.04.1992 at 04 h 50 min 23 s (with the magnitude $M = 6.1$ and hypocenter depth 12 km), and the second one on 28.06.1992 at 11 h 57 min 34 s (magnitude $M = 7.3$ and hypocenter depth is 1 km). The epicenters of the main shocks were approximately 30 km apart. The respective functions $g(t)$ and $\sigma(t)$ are shown in Fig. 4.

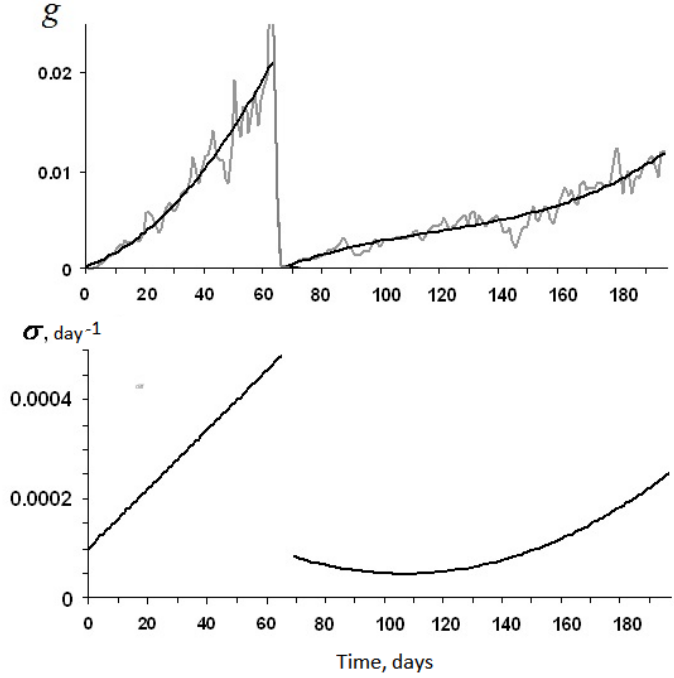

Fig. 4. Results of the analysis of the aftershocks shown in Fig. 3. Top panel: The function $g(t)$ before and after smoothing (the respective gray and solid lines). Bottom panel: The dependence of the deactivation coefficient $\sigma$ on time.

Thus, Eqn (2) and its solution (3) offer us the new possibility of analysis of the aftershocks data. There is an interesting perspective of introducing a new methodological technique: the classification of earthquake sources based on the shape of relaxation of aftershock sequences. For example, Figs 2 and 4 demonstrate two essentially different relaxation types. If we leave details aside, then in the first event we see a decreasing function $\sigma(t)$, and in the second two, increasing functions $\sigma(t)$. We assume that in the future the analysis of the function $\sigma(t)$ will allow us to introduce a more detail classification of the earthquake sources.

In conclusion of this section we will discuss briefly one more way of research begun by Fusakichi Omori over a hundred years ago. This way was found when we saw an analogy between variations in the state of rocks in the source and variations in the Earth's climate. Suppose there exists a quasi-equilibrium state $\bar{\sigma}(\varphi)$, where $\varphi(t)$ is a set of internal parameters of the source, which, generally speaking, depend on time. Then the relaxation theory of the deactivation of the source can be based on the equation

$$\frac{d\sigma}{dt} = \frac{\bar{\sigma}(\varphi) - \sigma}{\tau} + \xi(t), \qquad (7)$$

similar to that used to describe the average temperature of the earth's surface [Byalko, 2012]. Here $\tau$ is the characteristic relaxation time. The function $\xi(t)$ models some influences on the source. Integrating (6), we obtain

$$\sigma(t) = \left\{\sigma_0 + \int_0^t [\xi(t') + \tau^{-1}\overline{\sigma}(\varphi(t'))]\exp(t'/\tau)\,dt'\right\}\exp(-t/\tau). \qquad (8)$$

The choice of the perturbing function $\xi(t)$ requires special consideration. The internal disturbances (endogenous triggers) may be, for example, the round-the-world seismic echoes [Zavyalov et al., 2017; Zotov et al., 2018], and spheroidal oscillations of the Earth [Guglielmi et al., 2014]. Electromagnetic fields of natural or artificial origin can serve as external disturbances (exogenous triggers) [Buchachenko, 2014]. In each particular case one can choose a pulse, periodic or stochastic function $\xi(t)$. Specific interest is the simulation of space weather factors affecting the seismic activity by using $\xi(t)$.

We have indicated two promising directions for the study of aftershock physics. Another direction, which certainly deserves attention, is the study of the spatial-temporal distribution of aftershocks. Here, there are interesting opportunities for searches.

## 4. Conclusion

In memory of the outstanding scientist, we described briefly the law of Omori (1). Then we introduced an aftershocks equation (2), the solution of which is the Omori law. On this basis, we proposed original formula (3) that describes the evolution of aftershocks. This formula takes into account the non-stationarity of the earthquake source, which "cools down" after the main shock. The aftershocks equation allowed us to put the inverse problem of physics of the earthquake source.

We express our deep gratitude to B.I. Klain, A.S. Potapov, and O.D. Zotov for numerous discussions of the problem. This work was partially supported by the Program 28 of the Presidium of RAS, RFBR project # 18-05-00096, Russian Governmental assignment programs # 0144-2014-0097 and # 0144-2014-00103.

## References

Buchachenko A.L. Magnetoplasticity and the physics of earthquakes. Can a catastrophe be prevented? // Physics–Uspekhi. 2014. V. 57. No. 1. P. 92–98.

Byalko A.V. Relaxation theory of climate // Physics–Uspekhi. 2012. V. 55. No. 1. P. 103–108.

Costa L.S., Lenzi E.K., Mendes R.S., Ribeiro H.V. Extensive characterization of seismic laws in acoustic emissions of crumpled plastic sheets // EPL 114 59002 (2016) www.epljournal.org, doi: 10.1209/0295-5075/114/59002.

Davison Ch. Fusakichi Omori and his work on Earthquakes // Bulletin of the Seismic Society of America. December 01, 1924. V.14. P. 240–255.

Davison Ch. The founders of seismology // Cambridge: University Press, 1927) pp. 203–223.

Guglielmi A.V. Interpretation of the Omori law // arXiv:1604.07017. 24 Apr 2016.

Guglielmi A.V. Omori law: A note on the history of geophysics // Physics–Uspekhi. 2017. V. 60 (3). P. 319 – 324.

Guglielmi A.V., Zotov O.D., Zavyalov A.D. The aftershock dynamics of the Sumatra–Andaman earthquake // Izv. Phys. Solid Earth. 2014. V. 50. No. 1. P. 64–72.

Gutenberg B., Richter C.F. Frequency of earthquakes in California // Bull. Seismol. Soc. Am. 1944. V. 34 (4). P. 185–188.

Hirano R. Investigation of aftershocks of the great Kanto earthquake at Kumagaya // Kishoshushi. Ser. 2. 1924. V. 2. P. 77–83.

Omori F. On the aftershocks of earthquake // J. Coll. Sci. Imp. Univ. Tokyo. 1894. V. 7. P. 111–200.

Utsu T. A statistical study on the occurrence of aftershocks // Geophys. Mag. 1961. V. 30. P. 521–605.

Utsu T., Ogata Y., Matsu'ura R.S. The centenary of the Omori formula for a decay law of aftershock activity // J. Phys. Earth. 1995. V. 43, P. 1-33.

Zavyalov A., Zotov O., Guglielmi A., Lavrov I. Round-the-world seismic echo effect in aftershock sequences of strong earthquakes: a statistical analysis // Join IAG-IASPEI General Assembly. Kobe, Japan. July 30 – August 5, 2017. IASPEI Symposium / IASPEI03. Earthquake Generation Process / S09. Open session: Earthquake generation process – physics, modeling and monitoring for forecast. (invited report). https://confit.atlas.jp/guide/event/iagiaspei2017/subject/S09-1-01/classlist

Zotov O.D., Zavyalov A.D., Guglielmi A.V., Lavrov I.P. On the possible effect of round-the-world surface seismic waves in the dynamics of repeated shocks after strong earthquakes // Izv. Phys. Solid Earth. 2018. V. 54. No. 1. P. 178–191.